\documentclass[12pt]{article}
\usepackage[dvips]{graphicx}
\usepackage[tight]{subfigure}

\def\be{\begin{equation}}
\def\ee{\end{equation}}
\def\bea{\begin{eqnarray}}
\def\eea{\end{eqnarray}}
\def\beaN{\begin{eqnarray*}}
\def\eeaN{\end{eqnarray*}}
\def\ed{\end{document}}
\def\bit{\begin{itemize}}
\def\eit{\end{itemize}}

\def\sig{\sigma}

\def\lam{\lambda}

\def\Del{\Delta}

\def\k{\kappa}
\def\alf{\alpha}

\def\BD{\Bar D}

\def\di{\partial}

\def\half{{\textstyle{1 \over 2}}}
\def\~{\tilde}
\def\lag{{\hat{\cal L}}}
\def\m{\label}
\def\l{\left}
\def\r{\right}

\def\goto{\rightarrow}
\def\Bar{\overline}
\def\const{\rm const}
\def\diag{\rm diag}

\begin{document}

\centerline{\bf ON CREATING MASS/MATTER BY EXTRA DIMENSIONS}
\centerline{\bf IN THE EINSTEIN-GAUSS-BONNET GRAVITY}

\smallskip

\centerline{\it { A.N.Petrov}} \centerline{\it Inter-University
Center for Astronomy and Astrophysics,} \centerline {\it Post Bag 4,
Ganeshkhind Pune 411 007, INDIA} \centerline{ and}

\centerline{\it Relativistic Astrophysics group, Sternberg
Astronomical institute,} \centerline {\it
 Universitetskii pr., 13, Moscow, 119992,
RUSSIA}

\centerline{ E-mail: anpetrov@rol.ru}
\smallskip
PACS numbers: 04.50+h; 11.30.-j

 \begin{abstract}

Kaluza-Klein (KK) black hole solutions in the Einstein-Gauss-Bonnet
(EGB) gravity in $D$ dimensions obtained in the current series of
the works by Maeda, Dadhich and Molina are examined. Interpreting
their solutions, the authors claim that the mass/matter is created
by the extra dimensions. To support this claim, one needs to show
that such objects have classically defined masses. We calculate the
mass and mass flux for 3D KK black holes in 6D EGB gravity whose
properties are sufficiently physically interesting. Superpotentials
for arbitrary types of perturbations on arbitrary curved
backgrounds, recently obtained by the author, are used, and
acceptable mass and mass flux are obtained. A possibility of
considering the KK created matter as dark matter in the Universe is
discussed.
 \end{abstract}

\section{Introduction}
\m{Introduction}

We study new exact solutions in the Einstein-Gauss-Bonnet (EGB)
gravity in $D$ dimensions, which are $d$-dimensional Kaluza-Klein
(KK) black holes (BHs) with $(D-d)$-dimensional submanifold,
presented recently in \cite{MaedaDadhich1} - \cite{MaedaDadhich3} by
Maeda, Dadhich and Molina. The authors treat them as a classical
example of creating matter by curvature. The idea of such a kind is
not new. Thus, to make inflation possible, a pioneer proposal was
advanced by Starobinsky \cite{Star1} that a high-energy density
state was achieved by curved space corrections. Many other problems
of modern cosmology may be solved in the framework of
multidimensional gravity using high-order curvature invariants of KK
type spacetimes, see, e.g., \cite{Bronnikov} and references there
in.

To support the claim on creating `matter without matter', it is
necessary to calculate the mass and the mass flux by classical
methods. It is the main goal of the present paper. Here, we
concentrate on 3D BHs in 6D EGB gravity \cite{MaedaDadhich3}. These
toy objects are rich enough in physical properties, e.g., they can
have a radiative regime. For calculations we use the conservation
laws developed by us in \cite{Petrov2005b} - \cite{Petrov2009},
where in the framework of EGB gravity, superpotentials
(antisymmetric tensor densities) for arbitrary types of
perturbations on arbitrary curved backgrounds have been constructed.
Three important types of superpotentials \cite{Petrov2009} are used,
those based on (i) N{\oe}ther's canonical theorem, (ii) Belinfante's
symmetrization rule and (iii) a field-theoretical derivation.

The paper is organized as follows. In section \ref{3DBHs}, we
outline the solutions obtained in \cite{MaedaDadhich1} -
\cite{MaedaDadhich3} and describe necessary properties of the 3D
objects in 6D EGB gravity. In particular, in a natural way, we
define a spacetime where a BH is placed. It can be considered as a
possible background against which perturbations are studied. In
section \ref{mass+flux}, in the preliminaries, the main notions and
properties of the applied formalism are presented. Then we study the
objects themselves: (a) as vacuum 6D solutions; (b) as 3D KK
solutions with a `matter' created by extra dimensions. Calculating
the mass and the mass flux we support the second viewpoint. In
section \ref{Remarks}, we discuss (a) an ambiguity in the canonical
approach related to a divergence in the Lagrangian; (b) a
possibility of applying the KK BH solutions in cosmology. The
Appendix presents explicit general expressions for all three types
of superpotentials in EGB gravity.

\section{Kaluza-Klein 3D black holes}
 \m{3DBHs}
\setcounter{equation}{0}

We consider the action of the EGB gravity in the form:
 \be
 S  = -\frac{1}{2\k_D} \int d^D x\lag_{EGB} = -\frac{1}{2\k_D}
 \int d^D x\sqrt{-g} \l[R - 2\Lambda_0 +
 \alpha\underbrace{\l(R^2_{\mu\nu\rho\sig} - 4 R^2_{\mu\nu} +
 R^2\r)}_{\mbox{$L_{GB}$}}\r]
 \,
 \m{EGBaction}
 \ee
where $\alpha >0$. Here and below, curvature tensor
$R^{\mu}{}_{\nu\rho\sig}$, Ricci tensor $R_{\mu\nu}$ and scalar
curvature $R$ are related to the dynamic metric $g_{\mu\nu}$; a
`hat' means densities of the +1, e.g., $\hat g^{\mu\nu} =
\sqrt{-g}g^{\mu\nu}$; $({,\alf}) \equiv \di_{\alf}$ means ordinary
derivatives; the subscripts `${}_{E}$' and `${}_{GB}$' are related
to the Einstein and the Gauss-Bonnet parts in (\ref{EGBaction}).

The main assumption in \cite{MaedaDadhich1} - \cite{MaedaDadhich3}
is that the spacetime is locally homeomorphic to ${\cal M}^d
\times{\cal K}^{D-d} $ with the metric $g_{\mu\nu} ={\diag}
(g_{AB},r^2_0\gamma_{ab})$, $A,B = 0,\cdots,d-1;~a,b=d,\cdots,D-1$.
Thus, $g_{AB}$ is an arbitrary Lorentzian metric on ${\cal M}^d$,
$\gamma_{ab}$ is the unit metric on the $(D-d)$-dimensional space of
constant curvature ${\cal K}^{D-d}$ with $k=0,\,\pm 1$. Factor $r_0$
is a small scale of extra dimensions compactified by appropriate
identifications. The gravitational equations corresponding to the
EGB gravity action (\ref{EGBaction}) have the form:
 \be
 {\cal G}^\mu{}_\nu\equiv  G^\mu{}_\nu+\alf  H^\mu{}_\nu +
 \delta^\mu{}_\nu\Lambda_0=0\,,
  \m{EGBeqs}
 \ee
 where the Einstein tensor $G^\mu{}_\nu$ and $\delta^\mu{}_\nu$
 correspond to the Einstein part and $H^\mu{}_\nu$ corresponds to
 the GB part in (\ref{EGBaction}).
After all assumptions their decomposition is as follows:
 \bea
{\cal G}^A{}_B&\equiv& \l[1 + \frac{2k\alf}{r_0^2}(D-d)(D-d-1)
\r]{}_{(d)}{G}^A{}_B + \alf\, {}_{(d)}\!{H}^A{}_B \nonumber\\ &+&
\l[\Lambda_0 - \frac{k}{2r^2_0 }(D-d)(D-d-1)\l(1 +
\frac{k\alf}{r_0^2}(D-d-2)(D-d-3)\r)\r]{\delta}^A{}_B=0\,; \m{d}\\
{\cal G}^a{}_b &\equiv& {\delta}^a{}_b \l\{- \frac{{}_{(d)}\!{R}}{2}
+ \Lambda_0 - \frac{k}{2r^2_0 }(D-d-1)(D-d-2)- \alf
\l[\frac{k}{r^2_0 }(D-d-1)(D-d-2)\!\times
\r.\r.\nonumber\\&{}&\times\!\! \l.\l. \l({}_{(d)}\!{R} +
\frac{k}{2r^2_0 }(D-d-3)(D-d-4)\r) +
\frac{{}_{(d)}\!{L}_{GB}}{2}\r]\r\}=0 \m{D-d}
 \eea
where the subscript `${}_{(d)}$' means that a quantity is
constructed with the use of $g_{AB}$ only. As a result, one can see
that (\ref{d}) is a tensorial equation on ${\cal M}^d$, whereas
(\ref{D-d}) is a constraint for it. However to obtain more
interesting solutions one has to consider a special case that the
quantity ${\cal G}^A{}_B$ disappears {\em identically}. This is
possible for $d \le 4$ only because then ${}_{(d)}\!H_{\mu\nu}
\equiv 0$. Next, constants are chosen so as to suppress the
coefficients in (\ref{d}), which is possible if $D\ge d+2$, $k=-1$
and $\Lambda_0 <0$. Taking into account all the above, there remains
a single governing equation, the scalar equation (\ref{D-d}) on
${\cal M}^d$.

Here, we consider the solutions for $D=6$ and $d=3$ presented in
\cite{MaedaDadhich3}. A suitable set of constraints for the
constants is $r^2_0 = 12\alf =-3/\Lambda_0$. Then, the left hand
side of (\ref{d}) disappears identically. Keeping in mind that
${}_{(3)}\!{L}_{GB}\equiv 0$, one simplifies (\ref{D-d}) to obtain
 \be
{}_{(d)}\!{R}=2\Lambda_0\,,
 \m{R=2L}
 \ee
to which the static solution $g_{AB}(r)$ has been found:
 \be
 ds^2 = - fdt^2 +f^{-1}dr^2 +r^2d\phi\,,\qquad f\equiv r^2/l^2 +q/r -\mu
 \,.
 \m{KK-rmetric}
 \ee
Here, $\mu$ and $q$ are integration constants, and $l^2 \equiv
-3/\Lambda_0$. The Einstein tensor components for the solution
(\ref{KK-rmetric}) are
 \be
G^0_0 = G^1_1 =1/l^2
 -q/2r^3,~~ G^2_2 = 1/l^2 +q/r^3\,.
 \m{ET-r}
 \ee
As a space of a constant curvature, $(D-d=3)$-sector is completely
presented by its scalar curvature:
 \be
{}_{(D-d)}\!{R}= 6k/r_0^2=2\Lambda_0 = -1/2\alpha\,.
 \m{(D-d)R=2L}
 \ee

For comparison we consider the BTZ BH \cite{BTZ}. Its metric is
presented in the form
 \be
 ds^2 = - fdt^2 +f^{-1}dr^2 +r^2d\phi\,,\qquad f\equiv -r^2 \Lambda_0-\mu \,,
 \m{BTZmetric}
 \ee
which is a solution to the 3D {\em pure Einstein} equations. The
horizon radius $r_+$ of the BH is defined as $r^2_+ =
-\mu/\Lambda_0$, thus $r_+$ (and consequently a BH itself)
disappears for vanishing $\mu$. Therefore the integration constant
$\mu$ can be called the mass parameter. For $\mu \goto 0$, the
so-called {\em real vacuum} related to the BH (in another word, a
spacetime where a BH is placed) is defined by (\ref{BTZmetric}) with
$\Bar f =-r^2\Lambda_0$. However, such a spacetime is not maximally
symmetric, unlike AdS one. The latter with $\Bar f = -r^2 \Lambda_0
+1$ is approached when $\mu = -1$. A difference between a real
vacuum and a maximally symmetric vacuum is usual in BH solutions of
modified metric theories (see, e.g., \cite{BHS,Cai}); the BTZ BH is
the simplest illustration.

The solution (\ref{KK-rmetric}) is more complicated than
(\ref{BTZmetric}), although one has clear analogies with the BTZ
case. Considering BH solutions for simulating dark matter (see a
discussion in section \ref{Remarks}) we are more interested in the
cases with a horizon.  In (\ref{KK-rmetric}), the equation for the
event horizon is $l^2 q + r_+(r_+^2 - l^2\mu)= 0$. It is again
natural to choose a mass parameter $\~\mu$ in such a way that the BH
horizon disappears under vanishing $\~\mu$. This gives $\~\mu = \mu
- q/r_+$ and $r_+^2=l^2\~\mu$ (compare with the BTZ case), and
consequently $\~\mu>0$.  Then a real vacuum is defined by
(\ref{KK-rmetric}) with $\Bar f \equiv r^2/l^2 +q/r - q/r_+$, it is
again not maximally symmetric. The maximally symmetric AdS vacuum is
defined by (\ref{KK-rmetric}) with $\Bar f \equiv r^2/l^2 +1$. For
the latter, parameter $q$ is considered entirely as a perturbation
together with $\mu +1$. For $\~\mu\le 0$ a horizon does not exist,
this takes place, when $\mu>0$ with $q>2 l\l(\mu/3\r)^{3/2}$ or
$\mu\leq 0$ with $q\geq 0$.

The scalar equation (\ref{R=2L}) is also satisfied by the radiative
Vaidya metric $g_{AB}(v,r)$:
 \be
 ds^2 = - fdv^2 +2dvdr +r^2d\phi\,,\qquad f\equiv r^2/l^2  + q(v)/r -\mu(v)\,
 \m{KK-vmetric}
 \ee
where $\mu(v)$ and $q(v)$ now depend on the retarded/advanced time
$v$. Keeping in mind a possibility to form KK black holes
\cite{MaedaDadhich1} - \cite{MaedaDadhich3}, advanced time is more
interesting. Then (\ref{KK-vmetric}) can be connected with the
solution of the
 {\em form} (\ref{KK-rmetric}) by the transformation
 $
 dt = dv - dr/f(v,r)
 $.
After that, for {\em every constant} $v_0$, one can define {\em its
own} horizon (if it exists) and a corresponding real vacuum
analogously to the static case. The Einstein tensor components
corresponding to (\ref{KK-vmetric}) are
 \be
G^0_0 = G^1_1 =1/l^2
 -q/2r^3,~~G^1_0 = (\dot \mu r-\dot q)/2r^2,~~ G^2_2 = 1/l^2
 +q/r^3\,,
 \m{ET-v}
 \ee
where dot means $\di/\di v$. The scalar curvature of
$(D-d=3)$-sector is expressed again by (\ref{(D-d)R=2L}).

Considering (\ref{KK-rmetric}) and (\ref{KK-vmetric}) as solutions
 to the Einstein 3D equations on ${\cal M}^3$ (or, the same, EGB
equations because in (\ref{EGBeqs}) one has
${}_{(3)}\!H_{\mu\nu}\equiv 0$), one concludes that they are not
vacuum equations with a redefined cosmological
 constant $\Lambda  = \Lambda_0/3 =-1/l^2$. Indeed, both (\ref{ET-r}) and
 (\ref{ET-v}) show that a `matter' source ${\cal T}_{AB}$ with
 zero trace ${\cal T}^A{}_A = 0$ should exist, and the Einstein
 equations corresponding to (\ref{R=2L}) could be
 rewritten as
 \be
 {}_{(3)}\!{R}_{AB}-\half g_{AB}{}_{(3)}\!{R} +g_{AB}\Lambda =
\k_3 {\cal T}_{AB}\,.
 \m{EEwithTAU}
 \ee
A natural treating in \cite{MaedaDadhich1} - \cite{MaedaDadhich3} is
that ${\cal T}_{AB}$ is created by the compact extra dimensions.

\section{The mass and the mass flux for 3D black holes}
 \m{mass+flux}
\setcounter{equation}{0}

\subsection{Preliminaries}

Our calculation is based on differential conservation laws for
perturbations in a given background spacetime in the form:
  \be
 \hat {\cal I}^\alf(\xi) = \di_\beta \hat {\cal I}^{\alf\beta} (\xi)\,
 \m{generalCLs}
  \ee
where $\xi^\alf$ is a displacement vector, $\hat {\cal I}^\alf$ is a
vector density (carrent) and $\hat {\cal I}^{\alf\beta}$ is an
antisymmetric tensor density (superpotential). Thus,
$\di_{\alf\beta}\hat {\cal I}^{\alf\beta} \equiv 0$ and
$\di_{\alf}\hat {\cal I}^{\alf} =0$. The current contains
energy-momentum of both matter and metric perturbations, whereas the
superpotential depends on metric perturbations only. Integrating
$\di_{\alf}\hat {\cal I}^{\alf} =0$ and using the Gauss theorem one
obtains the integral conserved charges in a generalized form:
 \be
{\cal P}(\xi) = \int_\Sigma d^{D-1} x\,\hat {\cal I}^0(\xi) =
\oint_{\di\Sigma} dS_i \,\hat {\cal I}^{0i}(\xi)\,
 \m{charges}
 \ee
where $\Sigma$ is a $(D-1)$-dimensional hypersurface $x^0 = \const$,
$\di\Sigma$ is its  $(D-2)$-dimensional boundary, the zero indices
correspond to time or lightlike coordinates, and small Latin indices
correspond to space coordinates. Since we consider spherically
symmetric systems, we need $01$-components of the superpotentials in
(\ref{charges}) only.

The formalism describes exact (not infinitesimal) perturbations in
general. This is achieved if one one solution (dynamical) is
considered as a perturbed system with respect to another
(background) solution of the same theory. Thus conserved quantities
are defined with respect to a fixed (thought as known) spacetime,
e.g., a mass of a perturbed system on a given background. A
background can be both vacuum and non-vacuum, and usually is to be
chosen to correspond with problems under consideration. The task of
the present paper is calculating a global mass of the KK BHs
presented above. It is more important the mass defined with respect
to a spacetime, in which BH is placed because then with vanishing
BH, one obtains a zero mass. Therefore, first of all a {\em real
vacuum} described in previous section is chosen as a natural
background. Although such backgrounds are curved and nonsymmetric,
the technique used is powerful. Besides, as interesting and
important backgrounds we consider the AdS space. For such kinds of
backgrounds, perturbations are not infinitesimal in general.
However, we need in appropriate asymptotic of superpotentials in
(\ref{charges}) only. As one can see below, the fall-off integrands
in (\ref{charges}) both at spatial and at null infinity turns out to
be sufficiently strong to allow surface integrals to converge and to
give reasonable results.

In the previous section, the bar meant a quantity related to a
spacetime where a BH is `placed'; here and below, without
contradictions the bar means a quantity related to a background
spacetime as a structure of the formalism. As a natural choice, for
the above described static and radiative solutions we use the
background metric in the same forms (\ref{KK-rmetric}) and
(\ref{KK-vmetric}), respectively, where $\Bar f=\Bar f(r)$ can be
arbitrary in general but should be static. For calculating the
global mass $M$ we use the {\em timelike} Killing vector
 \be
 \xi^\alf = (-1, {\bf 0})\,.
 \m{Killing}
 \ee
It has this {\em unique form} for the above two generalized types of
background metrics: the zero component in (\ref{Killing}) can be
both timelike and lightlike; ${\bf 0}$ includes  5 or 2 space
dimensions in a 6D or 3D derivation, respectively. The metrics of
the real vacuum and AdS space just belong to the aforementioned
two types of background metrics and consequently also have a
timelike Killing vector of the unique form (\ref{Killing}). Then,
since (\ref{Killing}) is used every time, we will not recall this
frequently.

 \subsection{The BTZ solution}
\m{SBTZS}

As an example, we calculate the mass of the BTZ BH \cite{BTZ} with
the metric (\ref{BTZmetric}). We take the {\em Einstein parts} of
each of the superpotentials (\ref{CanEGB}), (\ref{BelEGB}) and
(\ref{SymEGB-h}), and, keeping in mind a 3D consideration,
calculate their $01$-components
 \bea
 {}_E{\hat{\cal I}}^{01}_C
 &=& \frac{\sqrt{-\Bar g_3}}{2\k_3 r}(f-\Bar f)\l[\frac{r\Bar{f}'}{2\Bar
 f f}\l(f - \Bar f \r) - 1\r]\,,
 \m{Can-E}\\
 {}_E{\hat{\cal I}}^{01}_B
 &=& \frac{\sqrt{-\Bar g_3}}{2\k_3 r}(f-\Bar f)\l[\frac{r\Bar{f}'}{2\Bar
 f f}\l(3f + \Bar f \r) -\frac{r{f}'}{
 f^2}\l(f + \Bar f \r) - 1\r]\,,
 \m{Bel-E}\\
 {}_E{\hat{\cal I}}^{01}_S
 &=& -\frac{\sqrt{-\Bar g_3}}{2\k_3 r}(f-\Bar f)\frac{\Bar{f}}{
 f}\,,
 \m{Sym-E}
 \eea
where the prime means $\di/\di r$. Taking into account a background
with $\Bar f = -r^2 \Lambda_0$, for which $f-\Bar f=-\mu$, and
substituting (\ref{Can-E}) - (\ref{Sym-E})   into (\ref{charges}),
we obtain, as $r\goto \infty$, the unique result
 \be
 M = \oint_{r\goto\infty}{}_E\hat {\cal I}^{01}d\phi = \frac{\pi\mu}{\k_3}\,,
 \m{BTZmass}
 \ee
which is quite acceptable for the global mass of the BTZ BH (see,
e.g., \cite{BHinHD}). The canonical superpotential has already
been checked for calculating (\ref{BTZmass}) in
\cite{DerKatzOgushi}, for the other superpotentials the result
(\ref{BTZmass}) could be considered as a nice test. Using the AdS
background with $\Bar f = -r^2 \Lambda_0+1$ one obtains
$M=\pi(\mu+1)/\k_3$.

\subsection{The static KK solution}
\m{SKKS}

Now let us turn to (\ref{KK-rmetric}); since it is the solution of
the EGB theory one should try to calculate the mass with using the
{\em full} formulae (\ref{CanEGB}), (\ref{BelEGB}) and
(\ref{SymEGB-h}) for this theory. The full background metric is to
be chosen as $\Bar g_{\mu\nu}= \Bar g_{AB} \times r^2_0\gamma_{ab}$.
Many formulae below take place for arbitrary $\Bar f$ in
(\ref{KK-rmetric}), although in specific calculations we choose
$\Bar f \equiv r^2/l^2 +q/r - q/r_+$. Let us turn to the
$(D-2)$-dimensional surface integral (\ref{charges}). Really, the
distant surface is considered in $(d=3)$-dimensional spacetime only,
whereas the integral over the $(D-d=3)$-dimensional compact space
could be interpreted as a constant, which `normalizes' the 6D
Einstein constant $\k_6$ to the 3D one $\k_3$. Indeed, one has for
the global mass constructed by (\ref{charges}):
  \be
M =  \oint_{\di\Sigma} dx^{D-2}\,\sqrt{-\Bar g_D}\, {\cal I}^{01}_D
= \oint_{r\goto\infty} d\phi\sqrt{-\Bar g_d}\, {\cal
I}^{01}_D\,\oint_{r_0} dx^{D-d}\sqrt{-\Bar g_{D-d}}=
V_{r_0}\oint_{r\goto\infty} d\phi\sqrt{-\Bar g_d}\, {\cal I}^{01}_D.
 \m{charges=0}
 \ee
Thus, since ${\cal I}^{01}_D \sim 1/\k_6$ one could set $\k_3 =
\k_6/V_{r_0}$. At first we follow this prescription.

With our assumptions, we find out that the Einstein parts of the
$01$-components of the superpotentials (\ref{CanEGB}),
(\ref{BelEGB}) and (\ref{SymEGB-h}) for the solution
(\ref{KK-rmetric}) are described only by the $d$-sector. Therefore,
to calculate the Einstein parts, it is sufficient to use Eqs.
(\ref{Can-E}) - (\ref{Sym-E}), but only with $\sqrt{-\Bar g_3}/\k_3$
replaced by $\sqrt{-\Bar g_D}/\k_6$ . For all cases, in the natural
background, the Einstein part in (\ref{charges=0}) gives a result
corresponding to (\ref{BTZmass}):
 \be
M_E = \pi\~\mu V_{r_0}/\k_6\, .
 \m{ME}
 \ee

We now construct the GB $01$-components of the superpotentials
(\ref{CanEGB}), (\ref{BelEGB}) and (\ref{SymEGB-h}) for the solution
(\ref{KK-rmetric}). They consist of two parts. The first one is pure
$(d=3)$-dimensional:
 \bea
 {}_{(d)}{\hat{\cal I}}^{01}_C &\equiv& 0\, ,
  \m{Can-E'}\\
 {}_{(d)}{\hat{\cal I}}^{01}_B
 &=& \frac{\alf\sqrt{-\Bar g_D}}{\k_6 r^2}\frac{\Bar{f}}{f}(f-\Bar f)
(rf''-f')\,,
 \m{Bel-E'}\\
 {}_{(d)}{\hat{\cal I}}^{01}_S &\equiv & 0 \,
 \m{Sym-E'}
 \eea
  (for brevity we suppress the subscript `GB'). For $\Bar f \equiv r^2/l^2
+q/r - q/r_+$, the behavior of (\ref{Bel-E'}) as $r\goto \infty$ is
$\sim 1/r^3$, thus each of the variants (\ref{Can-E'}) -
(\ref{Sym-E'}) gives a zero contribution into the integral
(\ref{charges=0}). The other part of the GB $01$-components is
determined by the intersecting terms of the $(d=3)$-sector and the
scalar curvature of the $(D-d=3)$-sector (\ref{(D-d)R=2L}):
 \bea
 {}_{(D-d)}{\hat{\cal I}}^{01}_C &=& \frac{\sqrt{-\Bar g_D}}{4\k_6}
\l[(f-\Bar f)' -\frac{\Bar f'}{f\Bar f}(f-\Bar f)^2+ \frac{2(f-\Bar
f)}{r} \r] \,,
  \m{Can-E''}\\
 {}_{(D-d)}{\hat{\cal I}}^{01}_B
 &=& \frac{\sqrt{-\Bar g_D}}{2\k_6}\l[\frac{(f-\Bar f)^2}{2f\Bar{f}}(f+\Bar f)'
 +\frac{f^2-\Bar f^2}{f}\l(\frac{f'}{f} - \frac{\Bar f'}{\Bar f}\r)+ \frac{f-\Bar
f}{r} \r]\,,
 \m{Bel-E''}\\
 {}_{(D-d)}{\hat{\cal I}}^{01}_S &=&
 \frac{\sqrt{-\Bar g_D}}{2\k_6 r}(f-\Bar f)\frac{\Bar{f}}{
 f}\,,
 \m{Sym-E''}
 \eea
 where the subscript `${}_{(D-d)}$' means that a quantity is without
 pure `${}_{(d)}$'-terms.
We remark that both (\ref{Can-E'}) and (\ref{Can-E''}) are unique
for each of (\ref{divd1}) and (\ref{divd2}). The asymptotic of each
of (\ref{Can-E''}) - (\ref{Sym-E''}) at spatial infinity in the
natural background is $ \sim - \~\mu$, and their substitution into
(\ref{charges=0}) gives the unique result:
 \be
M_{GB} = - \pi\~\mu V_{r_0}/\k_6\, .
 \m{MGB}
 \ee
Thus, keeping in mind (\ref{ME}) one can see that the global mass
 defined in the natural background by the total integral (\ref{charges=0}) is
{\em zero} in all the three approaches. The same result is valid if
the AdS background with $\Bar f = r^2/l^2 +1$ is
chosen.\footnote{The zero result has been recently obtained for a
similar situation by other methods as well by R.G. Cai, L.M. Cao,
and N. Ohta, ``Black holes without mass and entropy in Lovelocj
gravity'', {\em Phys. Rev. D}, {\bf 81}, 024018; ({\em Preprint}
arXiv:0911.0245 [hep-th]).}

At least, this result could be anticipated for the field-theoretical
approach. Indeed, the superpotential (\ref{SymEGB-h}) can be
connected directly with the linearized equations \cite{Petrov2005b}.
Contracting the latter with $\xi^\alf$ in (\ref{Killing}), one
selects the $d$-sector only. However, under the present assumptions,
the tensor in (\ref{d}) is equal to zero identically, therefore its
linearization is equal to zero identically as well. This conclusion
is supported by combining the expressions (\ref{Sym-E}), with the
replacement $\sqrt{-\Bar g_3}/\k_3 \goto \sqrt{-\Bar g_D}/\k_6$,
(\ref{Sym-E'}) and (\ref{Sym-E''}), which leads to zero identically.
At the same time, the canonical and Belinfante corrected approaches
give a zero result only asymptotically.

Of course, the zero result cannot be acceptable. Analyzing
(\ref{KK-rmetric}), one can find out that, considering this system
from the point of view of the Newtonian-like limit in 3 dimensions
(see, e.g., \cite{BHinHD}), this system must have a total mass. Thus
one should conclude that a {\em vacuum} 6D interpretation
(\ref{EGBeqs}) with (\ref{charges=0}) is not successful. By this
argument, one should consider the 3D Einstein interpretation
(\ref{EEwithTAU}) with a {\em created `matter'}. Calculating the
global conserved quantity basing on (\ref{charges}), we can use only
the surface integral, whereas a source (maybe not determined
explicitly, as in (\ref{EEwithTAU})) is included into the current in
the volume integral. Thus, considering the solution
(\ref{KK-rmetric}), we can be restricted to only the {\em Einstein
parts} of each of the superpotentials (\ref{CanEGB}), (\ref{BelEGB})
and (\ref{SymEGB-h}) related to the {\em non-vacuum} equations
(\ref{EEwithTAU}). As a {\em full} background metric,  one must
again consider $\Bar g_{AB}$ in (\ref{KK-rmetric}) (without
$r^2_0\gamma_{ab}$); we choose $\Bar f = r^2/l^2 +q/r -q/r_+$ again
and use the Killing vector (\ref{Killing}). Then, since the
parameter $q$ describes a `created matter' in (\ref{EEwithTAU}),
such a background is not vacuum in 3 dimensions now. Nevertheless,
the meaning of the notion `real vacuum' is not changed, although it
could be called wider as a `real background' now. Also, the applied
formalism remains powerful in non-vacuum backgrounds, and the
structure of the superpotentials remains the same. Then again we use
(\ref{Can-E}) - (\ref{Sym-E}) and obtain the acceptable result of
the type (\ref{BTZmass}):
 \be
 M = \pi\~\mu/\k_3\, .
 \m{MKK}
 \ee
If AdS space with $\Bar f = r^2/l^2 +1$ is chosen as a background,
the mass of the system is $M = \pi(\mu+1)/\k_3$. Note that in both
cases the parameter $q$ makes no contribution.

\subsection{The radiative Vaidya KK solution}
\m{RKKS}

For the radiative solution (\ref{KK-vmetric}) we have carried out
calculations similar to those in Subsection \ref{SKKS}. Though, in
this case the lightlike $v$-coordinate is used instead of the time
$t$-coordinate.  We again calculate $01$-components for the
superpotentials, however, now $\Sigma$ in (\ref{charges}) is defined
as $x^0 = v ={\rm constant}$, and the mass calculation is related to
null infinity. In Eqs. (\ref{Ev}) - (\ref{Ev+}) below, an arbitrary
$\Bar f =\Bar f(r)$ is considered. However, now there is no sense to
connect a background (which must be static) with a horizon (which is
changed in time). Therefore, in specific calculations we consider
the AdS background with $\Bar f = r^2/l^2+1$ only.

We first derive out the Einstein parts of all superpotentials:
  \be
 {}_E{\hat{\cal I}}^{01}_C = {}_E{\hat{\cal I}}^{01}_B =  {}_E{\hat{\cal I}}^{01}_S
 = -\frac{\sqrt{-\Bar g_D}}{2\k_6 r}(f-\Bar f)\,
 \m{Ev}
 \ee
where $f=f(v,r)$, which looks surprisingly simple, see, e.g.,
(\ref{Can-E}) - (\ref{Sym-E}). The GB $01$-components of the
superpotentials (\ref{CanEGB}), (\ref{BelEGB}) and (\ref{SymEGB-h})
for the solution (\ref{KK-vmetric}) consist of two parts again.
The pure $(d=3)$-dimensional part is
 \bea
 \l.{}_{(d)}{\hat{\cal I}}^{01}_C\r|_{(A.3)}& = &\frac{\alf\sqrt{-\Bar g_D}}{\k_6 r^2}
 (f-\Bar f)(f'-rf'')\,,
  \m{Can-E'+}\\
  \l.{}_{(d)}{\hat{\cal I}}^{01}_C\r|_{(A.4)} &\equiv& 0\,,
  \m{Can-E'+a}\\
  {}_{(d)}{\hat{\cal I}}^{01}_B
 &=& \frac{\alf\sqrt{-\Bar g_D}}{\k_6 r^2}\l[(f-\Bar f)
(rf''-f') + 2\l(r\Bar f'+ \Bar f +2r\frac{\di}{\di v}\r)\l(r(f-\Bar
f)' \r)' \r],
 \m{Bel-E'+}\\
 {}_{(d)}{\hat{\cal I}}^{01}_S &\equiv & 0 \,.
 \m{Sym-E'+}
 \eea
For the AdS background one has as $r\goto \infty$: for
(\ref{Can-E'+}) $\sim 1/r^3$ and for (\ref{Bel-E'+}) $\sim 1/r^2$,
thus all (\ref{Can-E'+}) - (\ref{Sym-E'+}) again give a zero
contribution into the integral (\ref{charges=0}). As in the static
case, the other part of the GB $01$-components is determined by the
intersection terms of the $(d=3)$-sector and the scalar curvature of
the $(D-d=3)$-sector (\ref{(D-d)R=2L}):
  \be
 {}_{(D-d)}{\hat{\cal I}}^{01}_C = {}_{(D-d)}{\hat{\cal I}}^{01}_B =
 {}_{(D-d)}{\hat{\cal I}}^{01}_S
 = \frac{\sqrt{-\Bar g_D}}{2\k_6 r}(f-\Bar f)\,.
 \m{Ev+}
 \ee
One can see that these components precisely compensate the
components (\ref{Ev}). Thus, as in the previous subsection, the
global mass defined in 6 dimensions is zero. Then one should follow
the interpretation of the static case and reject the vacuum 6D
derivation (\ref{EGBeqs}) with (\ref{charges=0}) as unacceptable
one.

We again consider Eq. (\ref{EEwithTAU}) as a governing one.
Restricting ourselves to the $d$-sector only and repeating the steps
of Subsection \ref{SKKS}, we obtain in the AdS background $M=
\pi(\mu(v)+1)/\k_3$. This is in a correspondence with the static
case. The mass flux for the radiating metric (\ref{KK-vmetric}) is
obtained simply by differentiating with respect to $v$: $\dot M=
\pi\dot{\mu}(v)/\k_3$. Comparing with the known BMS flux derivation
\cite{BMS}, this looks acceptable.

Concluding the section we assert that since the KK BH objects have
classically defined global mass and flux, they bring `matter'
created by extra dimensions and a special structure of the objects
themselves. If we set $q=0$ and $q(v)=0$, then, at least in the
static case, ${\cal T}_{AB} =0$ in (\ref{EGBeqs}). However, this
does not influence on our assertion because in all the cases $q$ and
$q(v)$ do not contribute into the global mass. Thus, mass/matter is
created in a more wide sense than creating ${\cal T}_{AB}$ in
(\ref{EGBeqs}).

\section{Concluding remarks}
 \m{Remarks}
\setcounter{equation}{0}

We will first discuss a well-known ambiguity in the canonical
approach related to a choice of a divergence in the Lagrangian. We
consider this problem in \cite{Petrov2009} and do not make a
definite choice between \cite{DerKatzOgushi} (or (\ref{divd1})) and
\cite{KatzLivshits} (or (\ref{divd2})). Indeed, both choices give an
acceptable mass for the Schwarzschild-AdS BH tested in
\cite{Petrov2009}. Here, the study of KK objects also does not give
an answer because in all cases we have a unique result. However, in
\cite{KatzLivshits} arguments in favor (\ref{divd2}) are given. In
multitimendional GR, the Katz and Livshits superpotential
\cite{KatzLivshits} turns out {\em uniquely} the KBL superpotential
\cite{KBL}; in EGB gravity, their superpotential naturally transfers
into the KBL superpotential for $D=4$. This is in a correspondence
with the Olea arguments \cite{Olea} where GB terms in the Lagrangian
regularize conserved quantities even if $D \le 4$. Lastly, the
choice (\ref{divd2}) looks more preferable because (a) it is more
`symmetric' than (\ref{divd1}), (b) the canonical superpotential
with (\ref{Can-E'+a}) gives a zero global integral in 6 dimensions
identically, as in the field-theoretical approach.

Now we turn to cosmological problems. As well known, the properties
of dark energy and of dark matter are very weakly constrained by the
cosmological observable data, therefore their derivation remains
very uncertain. Thus a search for acceptable models describing the
cosmic ingredients is very important, it is carried out very
intensively, and even dramatically, see, e.g., the recent papers,
reviews \cite{Varun1} - \cite{Teo} and references there in.

As an example, in the recent paper \cite{Ha}, recalling the 't Hooft
ideas of 1985, so-called `quantum black holes' are discussed as
elementary particles playing the role of the dark matter particles.
The latter are assumed as weak interacting matter particles (WIMPs),
which can have desirable TeV energies (see the aforementioned
reviews). `Quantum black holes' can be presented just like WIMPs,
they can be stable and do not radiate in the Hawking-Bekenstein
regime, unlike usual black holes.

Our main results show that the solutions (\ref{KK-rmetric}) and
(\ref{KK-vmetric}) have a classically defined mass and mass flux.
This just presents a possibility for the KK BHs to be presented in
the regime of `quantum black holes'. Thus, the topic of the present
paper, as we think, could be related to the dark matter problems.
Concerning this, we remark the following. First, since the parameter
$q$ can describe additional (to gravity) interactions, its presence
can suppress the WIMP idea. Then one needs to set $q=0$, which is
permissible, as has been remarked above. Second, it is desirable to
have a positive mass for WIMP objects. We support this because, if a
BH exists, one has $\~\mu> 0$ that leads to $M > 0$. Third, basing
on the radiating regime, in \cite{MaedaDadhich1} -
\cite{MaedaDadhich3} a scenario of forming KK BHs in EGB gravity was
suggested. One could try to develop this scenario for various
epochs. Keeping in mind all that, in future studies we plan an
examination of more realistic models presented in
\cite{MaedaDadhich1} - \cite{MaedaDadhich3}: they are 4D KK objects
in 6 and more dimensions of EGB gravity.

\subsection*{Acknowledgments} The author thanks very much Naresh Dadhich,
Alexey Starobinsky, Joseph Katz, Nathalie Deruelle and Rong-Gen Cai
for fruitful discussions and useful comments and recommendations.
Also, the author expresses his gratitude to professors and
administration of IUCAA, where the work was mainly elaborated and
finalized, for nice hospitality. The work is supported by the grant
No. 09-02-01315-a of the Russian Foundation for Basic Research.

\appendix

\section{Superpotentials in the
EGB gravity}
 \m{EGBsuperpotentials}
\setcounter{equation}{0}

In this Appendix, we represent an explicit form of the three types
of superpotentials for perturbations in the EGB gravity
\cite{Petrov2009}. The background quantities: Christoffel symbols
$\Bar \Gamma^\sig_{\tau\rho}$, covariant derivatives $\Bar D_\alf$,
the Riemannian tensor $\Bar R^\sig{}_{\tau\rho\pi}$ and its
contractions are constructed on the basis of a background
 $D$-dimensional spacetime metric $\Bar g_{\mu\nu}$.
It is a known (fixed) solution of EGB gravity; the bar means that a
quantity is a background one. One can find a detail derivation in
\cite{Petrov2009}. We first present the superpotential in the
{\em canonical prescription}:
  \bea
 \hat {\cal I}^{\alf\beta}_{C} &=&  {}_{E}\hat {\cal I}^{\alf\beta}_{C}
+{}_{GB}\hat {\cal I}^{\alf\beta}_{C} =
 {\k}^{-1}\l({\hat g^{\rho[\alf}\Bar D_{\rho} \xi^{\beta]}} +
\hat g^{\rho[\alf}\Delta^{\beta]}_{\rho\sig}\xi^\sig-  \Bar
{D^{[\alf} \hat\xi^{\beta]}} + \xi^{[\alf}{}_{E}\hat
d^{\beta]}\r)\nonumber\\
&+& {}_{GB}{\hat \imath^{\alf\beta}_C} - {}_{GB}\Bar{\hat
\imath^{\alf\beta}_C} + \k^{-1}\xi^{[\alf}{}_{GB}\hat d^{\beta]}\,
 \m{CanEGB}
 \eea
 where $\Del^\alpha_{\mu\nu} \equiv \Gamma^\alpha_{\mu\nu} - \Bar
{\Gamma}^\alpha_{\mu\nu} = \half g^{\alf\rho}\l( \BD_\mu g_{\rho\nu}
+ \BD_\nu g_{\rho\mu} - \BD_\rho g_{\mu\nu}\r)$ and\footnote{The
expression (\ref{CanSupEGB}) is differed from the correspondent one
in \cite{Petrov2009}, where the mistake has been found.
Nevertheless, the main results and conclusions in \cite{Petrov2009}
are not changed; see Corrigendum: {\em Class. Quantum Grav.} {\bf
27} (2010) 069801 (2pp); {\em Preprint} arXiv:0905.3622 [gr-qc] .}
 \bea
{}_{GB}\hat \imath^{\alf\beta}_C = &- & \frac{2\alf\sqrt{-g}}{\k}
 \l\{\Delta^{\rho}_{\lam\sig}R_\rho{}^{\lam\alf\beta} +
4\Delta^{\rho}_{\lam\sig}g^{\lam[\alf}R^{\beta]}_\rho +
\Delta^{[\alf}_{\rho\sig}g^{\beta]\rho}R\r\}\xi^\sig
 \nonumber\\&-&
\frac{2\alf\sqrt{-g}}{\k}\l\{R_\sig{}^{\lam\alf\beta} +
 4 g^{\lam[\alf}R^{\beta]}_\sig + \delta_\sig^{[\alf}g^{\beta]\lam}R
 \r\}\BD_\lam \xi^\sig\,.
 \m{CanSupEGB}
 \eea
The vector density $\hat d^\lam = {}_{E}\hat d^\lam + {}_{GB}\hat
d^\lam $ could be defined as in \cite{DerKatzOgushi} or following
the prescription of \cite{KatzLivshits}:
  \bea
\hat d^\lam_1 &=& {2\sqrt{-g}} \Delta^{[\alf}_{\alf\beta}
g^{\lam]\beta}  +
 4 \alf\sqrt{-g}\l(
R_\sig{}^{\alf\beta\lam} - 4 R^{[\alf}_\sig g^{\lam]\beta}+
\delta^{[\alf}_\sig g^{\lam]\beta} R\r)\Delta^\sig_{\alf\beta}\,,
 \m{divd1}\\
 \hat d^\lam_2 &=&
{2\sqrt{-g}} \Delta^{[\alf}_{\alf\beta} g^{\lam]\beta}  +
 4 \alf\sqrt{-g}\l(
R_\sig{}^{\alf\beta\lam} - 2R^{[\alf}_\sig g^{\lam]\beta}-
2\delta^{[\alf}_\sig R^{\lam]\beta}+ \delta^{[\alf}_\sig
g^{\lam]\beta} R\r)\Delta^\sig_{\alf\beta}\,.
 \m{divd2}
 \eea
The Einstein part in (\ref{CanEGB}) is precisely the KBL
superpotential \cite{DerKatzOgushi,KBL}, which in 4D general
relativity (GR) for the Minkowski background in the Cartesian
coordinates and with the translation Killing vectors $\xi^\alf=
\delta^\alf_{(\beta)}$ is just the well-known Freud superpotential
\cite{Freud39}.

The {\em Belinfante corrected} superpotential in EGB gravity is
  \be
 \hat {\cal I}^{\alf\beta}_{B} =  {}_{E}\hat {\cal I}^{\alf\beta}_{B}
+{}_{GB}\hat {\cal I}^{\alf\beta}_{B}=
 {\k}^{-1}\l( \xi^{[\alf} \BD_\lam \hat l^{\beta]\lam}
-\BD^{[\alf}\hat l^{\beta]}_{\sig}\xi^\sig  +\hat
l^{\lam[\alf}\BD_\lam \xi^{\beta]}\r)
 +{}_{GB}{\hat
\imath^{\alf\beta}_{B}} - {}_{GB}\Bar{\hat \imath^{\alf\beta}_{B}}
 \m{BelEGB}
 \ee
 where $\hat l^{\alf\beta}= \hat g^{\alf\beta}- \Bar{\hat
g}^{\alf\beta}$ and
 \bea
{}_{GB}\hat \imath^{\alf\beta}_{B}  &= & {\alf\over
\k}\BD_\lam\l\{\hat R_\sig{}^{\lam\alf\beta} +
 4 g^{\lam[\alf}\hat R^{\beta]}_\sig
 +\l[2\hat R_\tau{}^{\rho\lam[\alf}
-2\hat R^{\rho\lam}{}_\tau{}^{[\alf} - 8\hat R^\lam_\tau
g^{\rho[\alf}\r.\r.
 \nonumber\\&+& \l.\l.4
\hat R^\rho_\tau g^{\lam[\alf} +4 g^{\rho\lam} \hat R^{[\alf}_\tau +
2\hat R\l( \delta^\lam_\tau g^{\rho[\alf}- \delta^\rho_\tau
g^{\lam[\alf}\r)\r]\Bar g^{\beta]\tau}\Bar g_{\rho\sig} \r\}\xi^\sig
 \nonumber\\&-&{2\alf\over \k}\l\{{\hat R}_\sig{}^{\lam\alf\beta} +
 4 {g^{\lam[\alf}\hat R^{\beta]}_\sig} + \delta_\sig^{[\alf}
 g^{\beta]\lam}\hat R
 \r\}
 \BD_\lam \xi^\sig\,.
 \m{BelSupEGB}
 \eea
The Einstein part, ${}_{E}\hat {\cal I}^{\alf\beta}_{B}$, being
constructed in arbitrary $D$ dimensions, has precisely the form of
the Belinfante corrected superpotential in 4D GR \cite{PK}. In the
Minkowski background in the Cartesian coordinates and with the
translation Killing vectors ${}_{E}\hat {\cal I}^{\alf\beta}_{B}$,
it transforms to the well-known Papapetrou superpotential
\cite{Papapetrou48}.

Lastly, the superpotential in the field-theoretical derivation in
EGB gravity is
   \bea
\hat {\cal I}_{S}^{\alf\beta} & =& {}_{E}\hat {\cal
I}_{S}^{\alf\beta} + {}_{GB}\hat {\cal I}_{S}^{\alf\beta} =
{\k}^{-1}
 \l( \xi_\nu
\Bar D^{[\alf}\hat h^{\beta]\nu}- \xi^{[\alf} \Bar D_\nu \hat
h^{\beta]\nu} + \xi^{[\alf} \Bar D^{\beta]}\hat h - \hat
h^{\nu[\alf}\Bar D_\nu\xi^{\beta]} +\half \hat h \Bar D^{[\alf}
\xi^{\beta]}\r)
 \nonumber\\& +& {{4\over 3}}\l(
 2\xi_\sig \BD_\lam  \hat N_{{GB}}^{\sig[\alf|\beta]\lam}   -
\hat N_{{GB}}^{\sig[\alf|\beta]\lam}
 \BD_\lam  \xi_\sig\r)\,.
 \m{SymEGB-h}
 \eea
where $\hat h_{\alf\beta}=  {\sqrt{-\Bar g}}(g_{\alf\beta}- \Bar
g_{\alf\beta})$ and
  \bea
&{}& \hat N^{\rho[\lam|\mu]\nu}_{GB}=
 \m{NNNantis}
\\& - & \frac{3\alpha\sqrt{-\Bar g}}{4\k}\l\{h^\sig_\sig\l[\Bar g^{\nu[\lam}\Bar
g^{\mu]\rho} \Bar R + 2\Bar g{}^{\rho[\lam}\Bar R{}^{\mu]\nu}- 2\Bar
g{}^{\nu[\lam}\Bar R{}^{\mu]\rho} -  \Bar R^{\rho\nu\lam\mu}\r] +
\l(\Bar g^{\rho[\lam}h^{\mu]\nu}- \Bar g^{\nu[\lam}h^{\mu]\rho} \r)
\Bar R\r.\nonumber\\
& + &2\l(h{}^{\nu[\lam}\Bar R{}^{\mu]\rho}- h{}^{\rho[\lam}\Bar
R{}^{\mu]\nu}\r)+ 2\l(h{}^{\sig[\lam}\Bar g^{\mu]\rho}\Bar
R{}^{\nu}_{\sig} - h{}^{\sig[\lam}\Bar g^{\mu]\nu}\Bar
R{}^{\rho}_{\sig}\r)+ 2\l(h{}^{\sig\rho}\Bar g^{\nu[\lam}\Bar
R{}^{\mu]}_{\sig} - h{}^{\sig\nu}\Bar g^{\rho[\lam}\Bar
R{}^{\mu]}_{\sig}\r)  \nonumber\\ &-& \l. 2 \Bar g^{\nu[\lam}\Bar
g^{\mu]\rho}h^\sig_\tau \Bar R_\sig^\tau
 + 4\l(\Bar
R{}_\sig{}^{[\lam\mu][\rho}h{}^{\nu]\sig}+ \Bar
R{}_\sig{}^{[\rho\nu][\lam}h{}^{\mu]\sig} \r)+2h_{\sig\tau}\l(\Bar
R{}^{\sig\nu\tau[\lam}\Bar g^{\mu]\rho} - \Bar
R{}^{\sig\rho\tau[\lam}\Bar g^{\mu]\nu}\r) \r\}\,.
  \nonumber
 \eea
One obtains from (\ref{SymEGB-h}) the Deser-Tekin superpotential
\cite{DT2} if one chooses the AdS background. Again, doing
simplifications in 4 dimensions as above, one obtains the Papapetrou
superpotential \cite{Papapetrou48} (note, see \cite{Petrov2008},
that in 4D GR the Belinfante and field-theoretical approaches give
the same result). Under weaker restrictions, say, to AdS/dS
backgrounds in 4D GR, the superpotential (\ref{SymEGB-h}) goes to
the Abbott-Deser expression \cite{AbbottDeser82}.

\ed